\documentclass{aa}  
 \usepackage{xcolor} 
\usepackage{graphicx}
\usepackage{txfonts}
\usepackage{natbib}
\bibpunct{(}{)}{;}{a}{}{,}
\begin{document}

   \title{Pyramid wavefront sensor optical gains compensation using a convolutional model}

   \author{V. Chambouleyron
          \inst{1}\fnmsep\inst{2}
          \and
          O. Fauvarque\inst{1}\fnmsep\inst{3}
          \and
          P. Janin-Potiron\inst{2}\fnmsep\inst{1}
          \and
          C. Correia\inst{4}
          \and
          J-F. Sauvage\inst{2}\fnmsep\inst{1}
          \and
          N. Schwartz\inst{5}
          \and
          B. Neichel\inst{1}
          \and
          T. Fusco\inst{2}\fnmsep\inst{1}
          }

   \institute{Aix Marseille Univ, CNRS, CNES, LAM, Marseille, France\\
              \email{vincent.chambouleyron@lam.fr}
         \and
             ONERA The French Aerospace Laboratory, F-92322 Châtillon, France
          \and
          IFREMER, Laboratoire Detection, Capteurs et Mesures (LDCM), Centre Bretagne, ZI de la Pointe du Diable, CS 10070, 29280, Plouzane, France
          \and
          W. M. Keck Observatory, 65 - 1120 Mamalahoa Hwy., Kamuela, HI 96743, USA
          \and
          UK Astronomy Technology Centre, Blackford Hill, Edinburgh EH9 3HJ, United Kingdom  
             }

 
  \abstract
   {Extremely Large Telescopes have overwhelmingly opted for the Pyramid wavefront sensor (PyWFS) over the more widely used Shack-Hartmann WaveFront Sensor (SHWFS) to perform their Single Conjugate Adaptive Optics (SCAO) mode. The PyWFS, a sensor based on Fourier filtering, has proven to be highly successful in many astronomy applications. However, it exhibits non-linearity behaviors that lead to a reduction of its sensitivity when working with non-zero residual wavefronts. This so-called Optical Gains (OG) effect, degrades the close loop performance of SCAO systems and prevents accurate correction of Non-Common Path Aberrations (NCPA).}
   {In this paper, we aim at computing the OG using a fast and agile strategy in order to control the PyWFS measurements in adaptive optics closed loop systems.}
   {Using a novel theoretical description of the PyFWS, which is based on a convolutional model, we are able to analytically predict the behavior of the PyWFS in closed-loop operation. This model enables us to explore the impact of residual wavefront error on particular aspects such as sensitivity and associated OG. The proposed method relies on the knowledge of the residual wavefront statistics and enables automatic estimation of the current OG. End-to-End numerical simulations are used to validate our predictions and test the relevance of our approach.}
   {We demonstrate, using on non-invasive strategy, that our method provides an accurate estimation of the OG. The model itself only requires AO telemetry data to derive statistical information on atmospheric turbulence.  Furthermore, we show that by only using an estimation of the current Fried parameter $r_0$ and the basic system-level characteristics, OGs can be estimated with an accuracy of less than 10\%. Finally, we highlight the importance of OG estimation in the case of NCPA compensation. The proposed method is applied to the PyWFS. However, it remains valid for any WFS based on Fourier filtering subject from OG variations.}  
   {}

   \maketitle

\section{Introduction}

\begin{figure}[!h]
    \centering
    \includegraphics[scale=0.9]{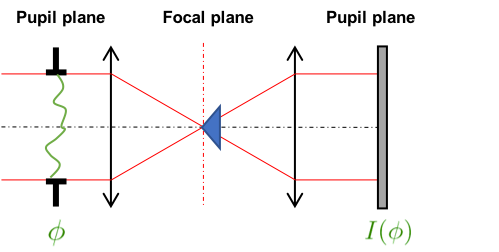}
    \caption{The PyWFS is a Fourier filtering wavefront sensor. A pyramidal mask is placed at a focal plane in order to achieve optical filtering. The output signal $I(\phi)$ show a relationship to the entrance phase $\phi$.}
    \label{fig:fourierFiltering}
\end{figure}

The Pyramid wavefront sensor (PyWFS) is an optical device used to perform wavefront sensing that was first proposed in 1996 \citep{raga}. Inspired by the Foucault knife test, the PyWFS is a pupil plane wavefront sensor performing optical Fourier filtering thanks to a glass pyramid located in the focal plane (see figure~\ref{fig:fourierFiltering}). This pyramid splits the electromagnetic (EM) field into four different beams, each producing four different filtered images of the entrance pupil. This filtering operation converts phase information at the entrance pupil into amplitude information at a pupil plane where a quadratic sensor is used to record the signal. The PyWFS usually includes an additional optical device called a modulation mirror. This mirror moves the Point-Spread Function (PSF) around the apex of the pyramid, which allows for an increase in the linearity range of the device at the expense of sensitivity.

\begin{figure}[!h]
    \centering
    \includegraphics[scale=0.7]{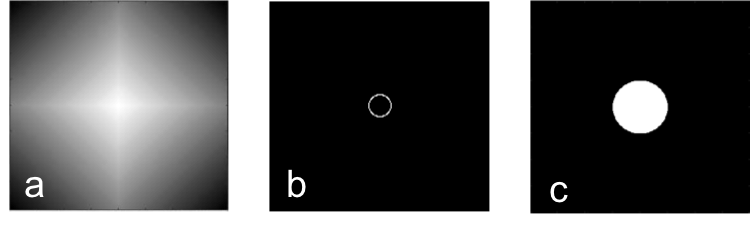}
    \caption{\textbf{a}: $arg(m)$ - shape of the pyramid mask. \textbf{b}: $w$ - modulation function. \textbf{c}: $\mathbb{I}_{p}$ - pupil shape.}
    \label{fig:convModel}
\end{figure}

The PyWFS displays higher sensitivity than the SHWFS and is therefore a key element for present and future AO systems. As an example, it will be used to perform the SCAO mode of all European Extremely Large Telescope (ELT) first light instruments \citep{harmoni,micado,metis}. Unfortunately, the PyWFS exhibits non-linear behaviours and the relationship between the produced signal and the incoming wavefront is not as straightforward as with the SHWFS.
The complexity and the limited knowledge on the nature of the PyWFS measurements led to extensive studies of this device \citep{verinaud,Guyon_2005,kkR,vicky} in order to analytically describe its linear response. However, it is possible to describe the PyWFS as a convolutional system that can be fully characterized by the knowledge of its impulse response, as it is widely done for many physical systems. The advantages of such a convolutional description are numerous: it allows for a fast numerical computation of the sensor's response to a given input phase and gives the frequency-dependent sensitivity through the transfer function of the system.
A first step in that direction was proposed by \citep{vicky}, but the model suffers from strong approximations (for example, the PyWFS is described as two rooftop masks and some terms are neglected to simplify calculations). To the best of our knowledge, the most complete study to date of the PyWFS as a convolutional system has been proposed by \citep{fauv}. In this model, the PyWFS is simply described by its three main properties: the shape of the pyramid mask $m$, the modulation function $w$, and the entrance pupil geometry $\mathbb{I}_{p}$ (see figure~\ref{fig:convModel}). According to this description, the impulse response of the system is given by the following equation:

\begin{equation}
    \textbf{IR} = 2\textbf{Im}(\bar{\widehat{m}}(\widehat{m}\star\widehat{w}\mathbb{I}_{p}))
    \label{eq:IR}
\end{equation}

where $\textbf{Im}$ is the imaginary part, $\ \widehat{}\ $ the Fourier transform operator and $\star$ the convolution symbol.

 Now that the PyWFS has captured the interest of AO scientists, one of its major limitation needs to be handled, namely its strong non-linear behaviour which leads to a spatial frequency-dependent loss of sensitivity during on-sky operations. This loss of sensitivity can be captured in a quantity called Optical Gains (OG) \citep{4k,vdeo}. Tracking the OGs during on-sky operations has therefore become one of the key priorities to fully control PyWFS measurements.

 OGs originating from non-linear behaviours have already been recognised in other WFSs, such as the quad-cells SHWFS \citep{Veran:00} or the Zernike WFS \citep{arthurZelda}. The main impact of OG in closed-loop operation is to introduce an error in the wavefront reconstruction. This error becomes predominant in the case of bad seeing conditions and/or when pointing at extended objects. There are new robust strategies available for on the fly optimisation of the loop gains in order to mitigate the reconstruction error impacted by OGs \citep{close}. However, these techniques do not give direct access to the actual  OG values (see section~\ref{NCPA}). In fact, knowledge of OG is essential for Non-Common Path Aberrations (NCPA) correction, which is emerging as a critical step in wavefront control for PyWFS based systems \citep{espositoNCPA}. The knowledge of OG is also a key issue in PSF reconstruction, where accurate analysis of loop telemetry data is paramount. The objective of this paper is to present a new strategy based on a physical description of the PyWFS to quickly and accurately compute the OGs, and that independently from the temporal loop gains.

In section~\ref{ogDef}, we present the definition of the OG and ways to better understand the physical nature of OG, which are generated by residual phases on the PyWFS. In section~\ref{convMod}, we then show that it is possible to use the convolutional model to accurately compute OG, provided some statistical information on the shape of the residual phases. Finally, in the last section of this paper, we demonstrate the superiority of our method for NCPA compensation.

\section{Definition of optical gains and application to PyWFS in presence of residual phases}
\label{ogDef}
\subsection{The interaction matrix as a linear model of the PyWFS}

The wavefront sensor can be described by a matrix that fully encodes the linear behaviour of the system. This so-called Interaction Matrix ($IM$) is computed through a calibration process by recording the slopes of the linear responses of the wavefront sensor to a set of incoming phases $\phi_{i}$. Combined, these wavefronts represent the basis of the phase space we want to control. For each mode, the slopes of the linear response $\delta I_{\text{calib}}(\phi_{i})$ can be computed through the following operation, often referred to as "push-pull": 
\begin{equation}
    \delta I_{\text{calib}}(\phi_{i}) = \frac{I_{\text{calib}}(a\phi_{i})-I_{\text{calib}}(-a\phi_{i})}{2a}
\end{equation}

where $I_{\text{calib}}$ is the recorded intensity on the wavefront sensor detector. A reference signal, corresponding to a flat wavefront in the pupil plane, is also subtracted from this value. In this paper, we use the full-frame definition for the PyWFS signal, however this work can easily and straightforwardly be applied the slope-like definition of the PyWFS measurements. In the previous equation, $a$ represents the amplitude of the mode used for calibration. $a$ should be as small as possible in order to stay within the linear regime of the sensor. But in reality, we want it to be large enough to ensure a satisfactory the signal-to-noise ratio, while at the same time staying within the linearity regime. This maximization of signal-to-noise ratio during calibration can be helped by using optimal calibration strategies, such as the Hadamard approach \citep{meimon}. The interaction matrix computed during the calibration process $IM_{\text{calib}}$ is then the concatenation the slopes recorded for all modes.
\begin{equation}
    IM_{\text{calib}} = (\delta I_{\text{calib}}(\phi_{1}),..., \delta I_{\text{calib}}(\phi_{i}),..., \delta I_{\text{calib}}(\phi_{N}))
\end{equation}

In the well-known inverse problems framework, this calibration step is actually a way to compute the \textbf{linear forward operator} of our system, associating the incoming wavefront with pyramid measurements. 

\begin{figure}[!h]
    \centering
    \includegraphics[scale=0.6]{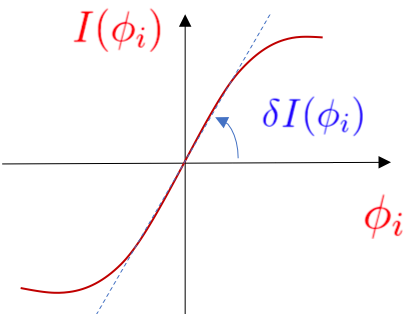}
    \caption{Sketch of the PyWFS response curve for a given mode $\phi_{i}$. The push-pull method around a null-phase consists in computing the slope of this curve for $a=0$.}
    \label{fig:linearityCurve}
\end{figure}

\subsection{Optical Gains: an offset between calibration regime and on-sky regime}

 $IM_{\text{calib}}$ is computed in a specific regime that we call the \textit{calibration regime}. The calibration is usually done using a point-like source around a flat wavefront (no reference phase) and for a given modulation radius.

 During operation - which we call the \textit{on-sky regime} - the wavefront sensor differs inevitably from the calibration regime, if nothing else because we can't reach the perfect diffraction limit of the telescope. Because of the non-linear nature of the PyWFS, this leads to a change in the behaviour of the sensor. It is possible to account for these non-linearities by considering the PyWFS as a sensor with a varying linear behaviour that depends on the current sensing regime. We therefore make the hypothesis that the sensor's behaviour in the on-sky regime can be described by an IM that we call $IM_{\text{onSky}}$. In that case, the linear behaviour has to be measured again to have an accurate description of the direct problem.
 
\begin{equation}
    \delta I_{\text{onSky}}(\phi_{i}) = \frac{I_{\text{onSky}}(a\phi_{i})-I_{\text{onSky}}(-a\phi_{i})}{2a}\
\end{equation}

 When the PyWFS is working around a non-null reference phase, we have the following relationship:
 
\begin{equation}
I_{\text{onSky}}(a\phi_{i}) = I_{\text{calib}}(a\phi_{i}+\phi_{res})
\end{equation}

because of the PyWFS's non-linear behaviour, we have $I_{\text{calib}}(a\phi_{i}+\phi_{res})\neq I_{\text{calib}}(a\phi_{i})+I_{\text{calib}}(\phi_{res})$ and therefore:

\begin{equation}
\begin{split}
\delta I_{\text{onSky}}(\phi_{i}) &=\frac{I_{\text{onSky}}(a\phi_{i})-I_{\text{onSky}}(-a\phi_{i})}{2a}\\
&=\frac{I_{\text{calib}}(a\phi_{i}+\phi_{res})-I_{\text{calib}}(-a\phi_{i}+\phi_{res})}{2a}\\
&\neq \delta I_{\text{calib}}(\phi_{i})
\end{split}
\end{equation}
which naturally leads to offsets between $IM_{\text{calib}}$ and  $IM_{\text{onSky}}$.\\

We define the \textbf{optical transfer matrix} $T_{\text{opt}}$ as the transfer matrix describing the offsets between the on-sky regime and the calibration regime. This matrix is a square matrix of size $N_{\text{modes}}$ x $N_{\text{modes}}$.
\begin{equation}
    IM_{\text{onSky}} = IM_{\text{calib}}.T_{\text{opt}}
    \label{eq:OGmatrix}
\end{equation}

In order to obtain the correct linear description of the sensor in a given sensing regime, we therefore need to adjust the interaction matrix computed during calibration by the optical transfer matrix.\\

From the equation above, we can write the exact definition of the optical transfer matrix:
\begin{equation}
    T_{\text{opt}} = IM_{\text{calib}}^{\dag}.IM_{\text{onSky}}
\end{equation}

\subsection{Diagonal approximation and OG definition in the PyWFS measurement space}

The \textbf{diagonal approximation} can strongly simplify the computation of $T_{\text{opt}}$. This approximation consists in  assuming that $T_{\text{opt}}$ is a diagonal matrix \citep{vdeo}, meaning there is no cross-talk between modes when we are switching from the calibration regime to the on-sky (or sensing) regime. In other words, the slope of the linear behaviour for each mode $\phi_{i}$ is increased or reduced by a scalar factor $G(\phi_{i})$ called the modal OG.\\

In the case of the diagonal approximation, we can define the modal OG $G(\phi_{i})$ without having to use the pseudo inverse $IM_{\text{calib}}^{\dagger}$ (which depends on the condition number): we propose the use of the scalar product $\langle \boldsymbol{\cdot}|\boldsymbol{\cdot}\rangle$ defined in the measurement space to compare $\delta I_{\text{onSky}}(\phi_{i})$ and $\delta I_{\text{calib}}(\phi_{i})$ for each mode $\phi_{i}$.

\begin{equation}
     G(\phi_{i})=\frac{\langle\delta I_{\text{onSky}}(\phi_{i})| \delta I_{\text{calib}}(\phi_{i})\rangle}{\langle\delta I_{\text{calib}}(\phi_{i})| \delta I_{\text{calib}}(\phi_{i})\rangle}
     \label{eq:OGwfsSpace}
\end{equation}

$\langle\delta I_{\text{onSky}}(\phi_{i})| \delta I_{\text{calib}}(\phi_{i})\rangle$ represents the projection of the measurement in the sensing regime onto the measurement in the calibration regime and $\langle\delta I_{\text{calib}}(\phi_{i})| \delta I_{\text{calib}}(\phi_{i})\rangle$ is a normalisation term. The definition of OG given here differs slightly from the ones previously given in the literature \citep{4k,vdeo}, and has the advantage of being independent of the reconstructor. This is a description in measurement space only. An equivalent formulation of equation~\ref{eq:OGwfsSpace} in terms of matrices is the following:

\begin{equation}
     G_{\text{opt}}=\frac{\text{diag}(^{\textbf{t}}IM_{\text{onSky}}.IM_{\text{calib}})}{\text{diag}(^{\textbf{t}}IM_{\text{calib}}.IM_{\text{calib}})}
     \label{eq:OGmatrix}
\end{equation}

where $G_{\text{opt}}$ is a vector containing all the $G(\phi_{i})$ for $i\in [1,N_{modes}]$.

\subsection{Impact of residual phases on the PyWFS impulse response}

The offset experienced by $IM_{\text{calib}}$ changes at each measurement because $\phi_{res}$ is a time-varying quantity. That is to say that $IM_{\text{onSky}}$ is changing at every iteration, depending on the content of $\phi_{res}$. Although it seems hard to determine the state of $IM_{\text{onSky}}$ at each instant, we can find a way to compute the \textbf{averaged state} of the sensing regime $<IM_{\text{onSky}}>_{t}$, which gathers $<\delta I_{\text{onSky}}(\phi_{i})>_{t}$ for each mode. 

\begin{equation}
 \begin{split}
    <IM_{\text{onSky}}>_{t} = (<\delta I_{\text{onSky}}(\phi_{1})>_{t},..., \\
    <\delta I_{\text{onSky}}(\phi_{i})>_{t},..., <\delta I_{\text{onSky}}(\phi_{N})>_{t})
    \label{eq:average}
\end{split}
\end{equation}

In this regard, we rely on the convolutional formalism of the PyWFS proposed by \cite{fauv}. Within the framework of this model, it is possible to compute an analytic function to take into account the impact of residual phases on PyWFS measurements. The sensing regime is then described by a PyWFS for which the modulation function (see equation \ref{eq:IR}) is changed according to this formula:
    \begin{equation}
    w \leftarrow w\star\widehat{e^{-\frac{1}{2}\text{D}_{\phi_{res}}}}
    \end{equation}
    
where $\text{D}_{\phi_{res}}$ is the residual phase structure function. This equation provides a fundamental insight into PyWFS measurements in the presence of residual phases. It was well-known that residual phases act as an extra modulation that lowers the pyramid's sensitivity. We are now able to quantify this loss: the impact depends on residual phases statistics through the structure function, and therefore through the Power Spectral Density (PSD). It is then possible to define its new impulse response in the averaged sensing regime assuming isotropy and stationarity of the residual phases:

\begin{equation}
    \textbf{IR}_{\textbf{onSky}} = 2\textbf{Im}(\bar{\widehat{m}}(\widehat{m}\star \widehat{w}\mathbb{I}_{p}e^{-\frac{1}{2}\text{D}_{\phi_{res}}}))
    \label{eq:IRsensing}
\end{equation}

The reader will notice that the modulation function is the only quantity affected here. This means that the impact of residual phases can be described as a collection of incoherent tip-tilt offsets during one measurement cycle. Changing from an apparently coherent offset to an incoherent offset, comes from the time averaging operation. 
This in fact is very well understood in the image formation field through the derivation of the atmospheric transfer function \cite{roddier81}. By averaging over time, we can derive an analytic formulation for the long-exposure seeing limited PSF, which cannot be fully described using a coherent phase aberration in the pupil plane.\\

In this section, we have presented a new measurement space based definition for the OG. We have also explained how they naturally emerge from PyWFS non-linearities when working with offsets between calibration and sensing regimes. In the following part, we propose a new method based on the convolutional model to perform a fast and accurate computation of the OG.

\section{A new strategy to compute PyWFS modal Optical Gains through the convolutive model}
\label{convMod}

\subsection{Convolutional formalism: a path to optical Gains computation}

In case of OG introduced by residual phase, the diagonal approximation ensures that the knowledge of the diagonal elements of $G_{\text{opt}}$ is sufficient to compute $IM_{\text{onSky}}$. The expression of  $G(\phi_{i})$ given equation~\ref{eq:OGwfsSpace} can be rewritten within the convolutional model using the impulse responses of the calibration regime and the sensing regime:

\begin{equation}
     G_{\textbf{conv}}(\phi_{i})=\frac{\langle\textbf{IR}_{\textbf{onSky}}\star\phi_{i}|\textbf{IR}_{\textbf{calib}}\star\phi_{i}\rangle}{\langle\textbf{IR}_{\textbf{calib}}\star\phi_{i}| \textbf{IR}_{\textbf{calib}}\star\phi_{i}\rangle}
     \label{eq:Gconv}
\end{equation}

We now have the means to compute the modal OG by knowing the following system parameters: the shape of the mask $m$, the modulation function $w$, the shape of the pupil $\mathbb{I}_{p}$, and the residual phase structure function $\text{D}_{\phi_{res}}$. 
In order to identify whether the convolutional model used here is sufficiently accurate to provide a good estimation of the modal OG - \textit{i.e.} whether  $G_{\textbf{conv}}(\phi_{i})$ is a good estimate of $G(\phi_{i})$ or not - we compared the predictions of the model with End-to-End simulations. The results of this study are presented in the next section.

\subsection{Convolutional model versus End-to-End simulations}

The End-to-End simulations are performed using the \textit{OOMAO} \textsc{Matlab} toolbox \citep{oomao}, considering a \textbf{8 m class telescope}. The resolution in the pupil diameter is 90 pixels across. We use a Karhunen-Loève basis composed of 400 modes to compute all our interaction matrices and OG. The wavefront sensing is done in the visible ($\lambda = 550\ nm$).

\subsubsection*{Sensitivity curves}

\begin{figure}[!h]
    \centering
    \includegraphics[scale=0.42]{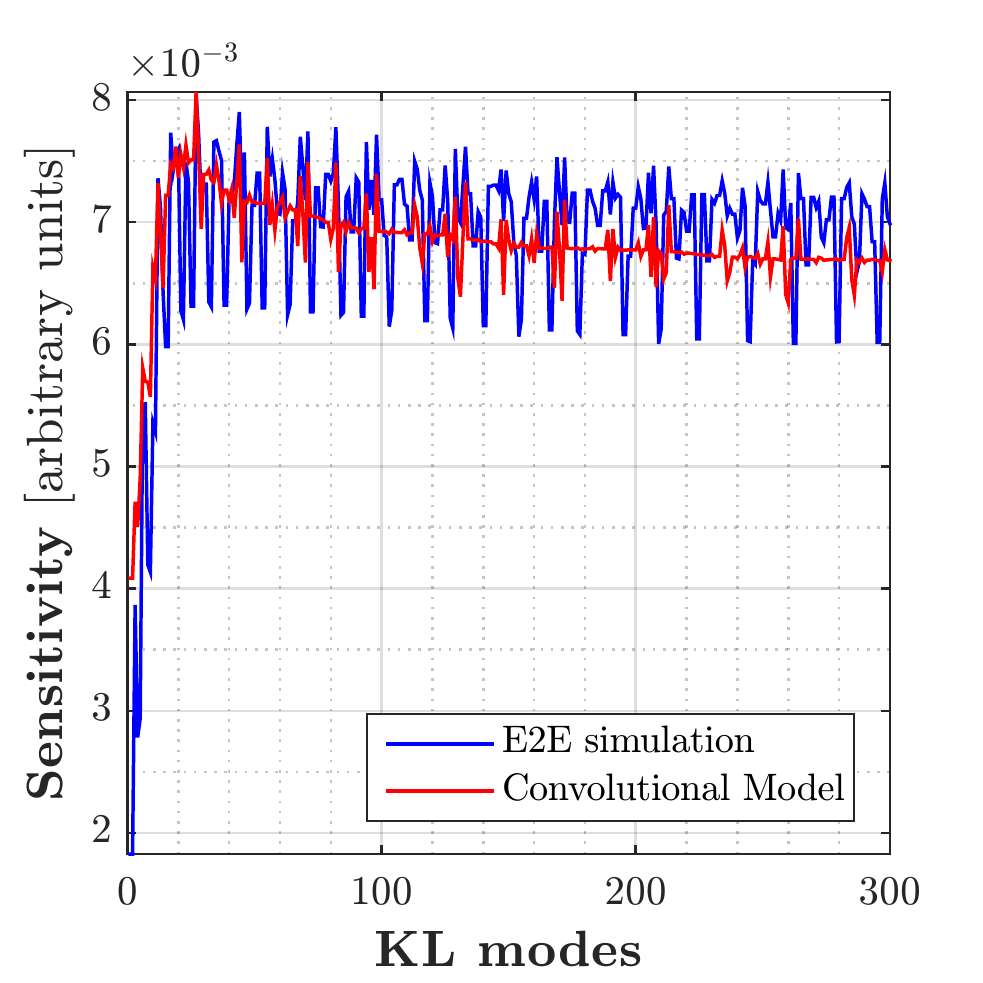}
    \includegraphics[scale=0.42]{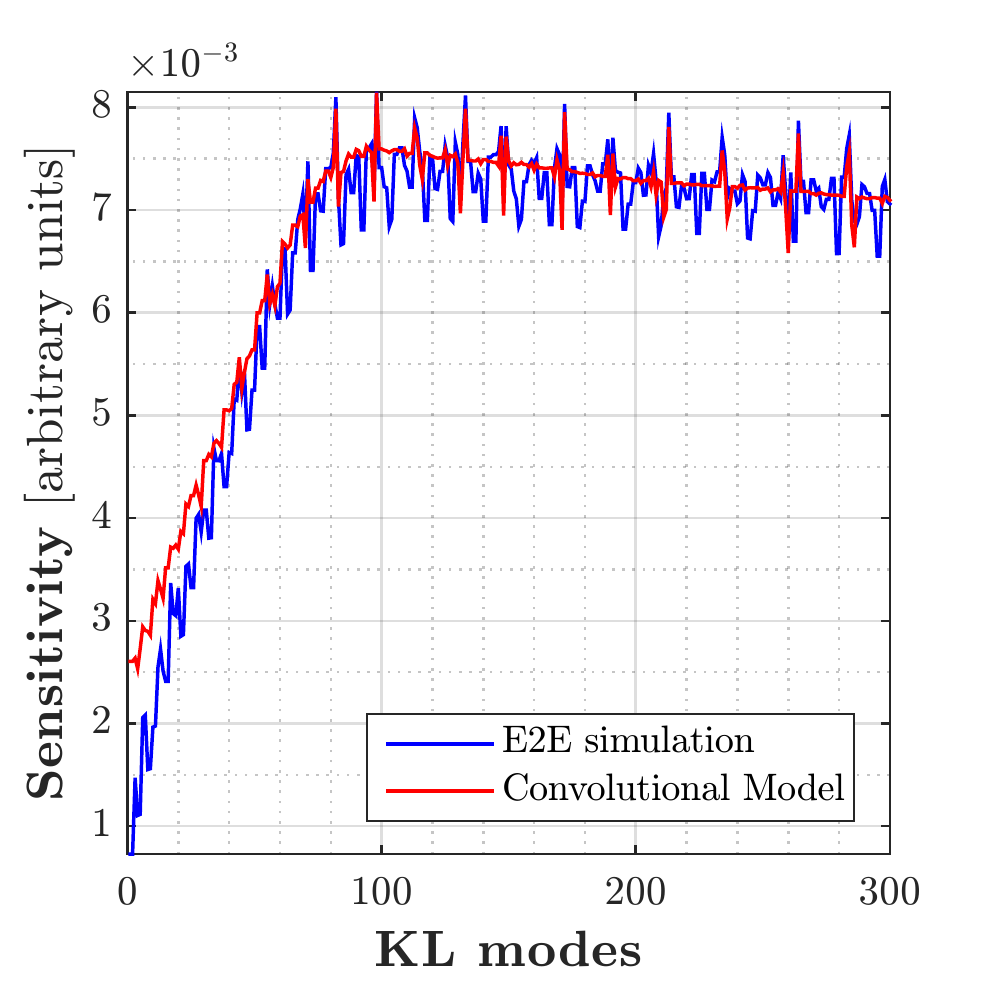}
    \caption{The well-known pyramid sensitivity curves. \textbf{Left}: Modulation radius $r_{mod} = 2\lambda/D$. \textbf{Right}: Modulation radius $r_{mod} = 5\lambda/D$.}
    \label{fig:sensitivity}
\end{figure}

We use the convolutional model to compute the well-known sensitivity curves of the PyWFS where the sensor behaves as a slope sensor for the frequencies lower than the modulation radius and as a phase sensor for the frequency above. For the chosen system configuration, we present figure~\ref{fig:sensitivity} results for two different modulation radii. We remind the reader that for each mode, the sensitivity is given by:

\begin{equation}
     s(\phi_{i})=||\delta I_{\text{calib}}(\phi_{i})||_{2}=\sqrt{\langle\delta I_{\text{calib}}(\phi_{i})| \delta I_{\text{calib}}(\phi_{i})\rangle}
\end{equation}

We note a small offset between the model and the end-to-end simulations for the low-order modes. This can be explained by the hypothesis of the sliding pupil used in the derivation of the convolutional model. This issue was presented in \cite{fauv}.

\subsubsection*{Modal Optical Gains}

We carry out the study by computing modal OG through End-to-End simulations in different system configurations. We then compare those to the ones predicted through the convolutional model. We suppose here that we know the turbulence statistics. In other words, we have access to the PSD (Power Spectral Density) or the structure function of the residual phases. We will focus on how to get this data in a practical way later in this paper. \\

\begin{figure}[!h]
    \centering
    \includegraphics[scale=0.42]{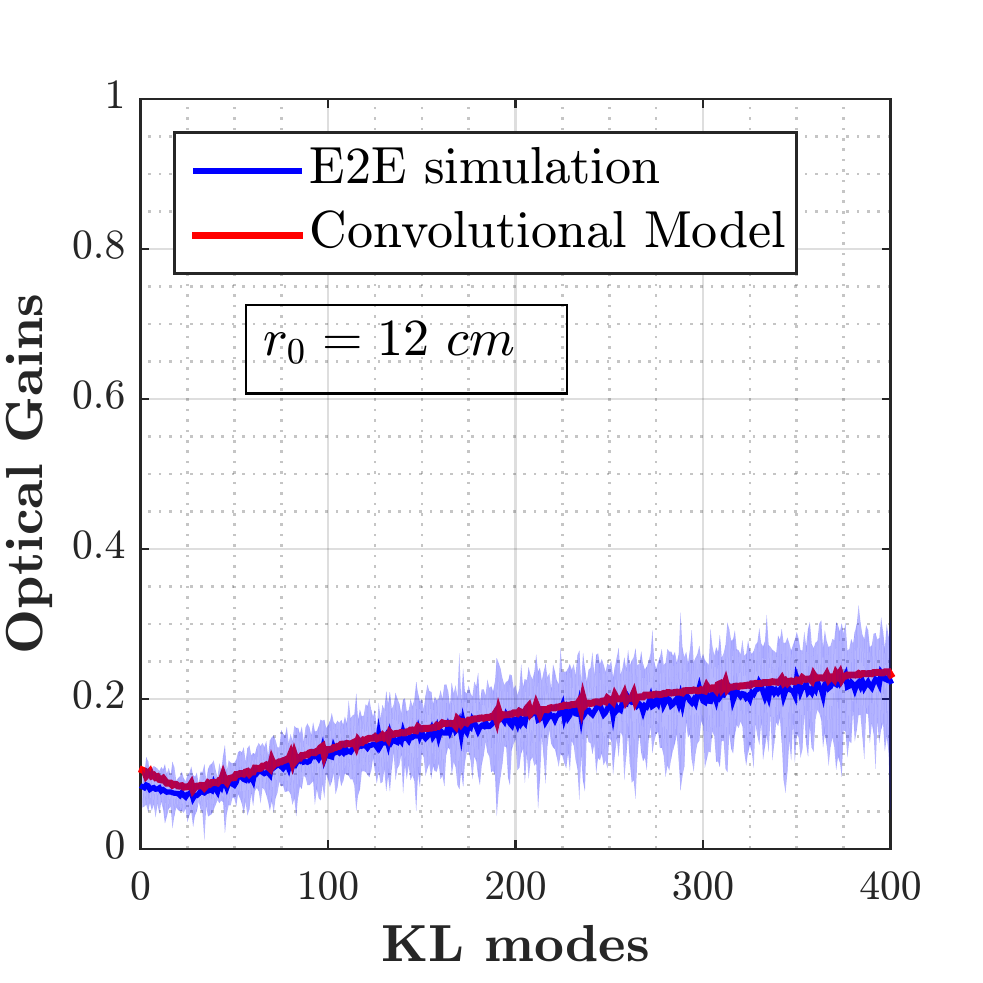}
    \includegraphics[scale=0.42]{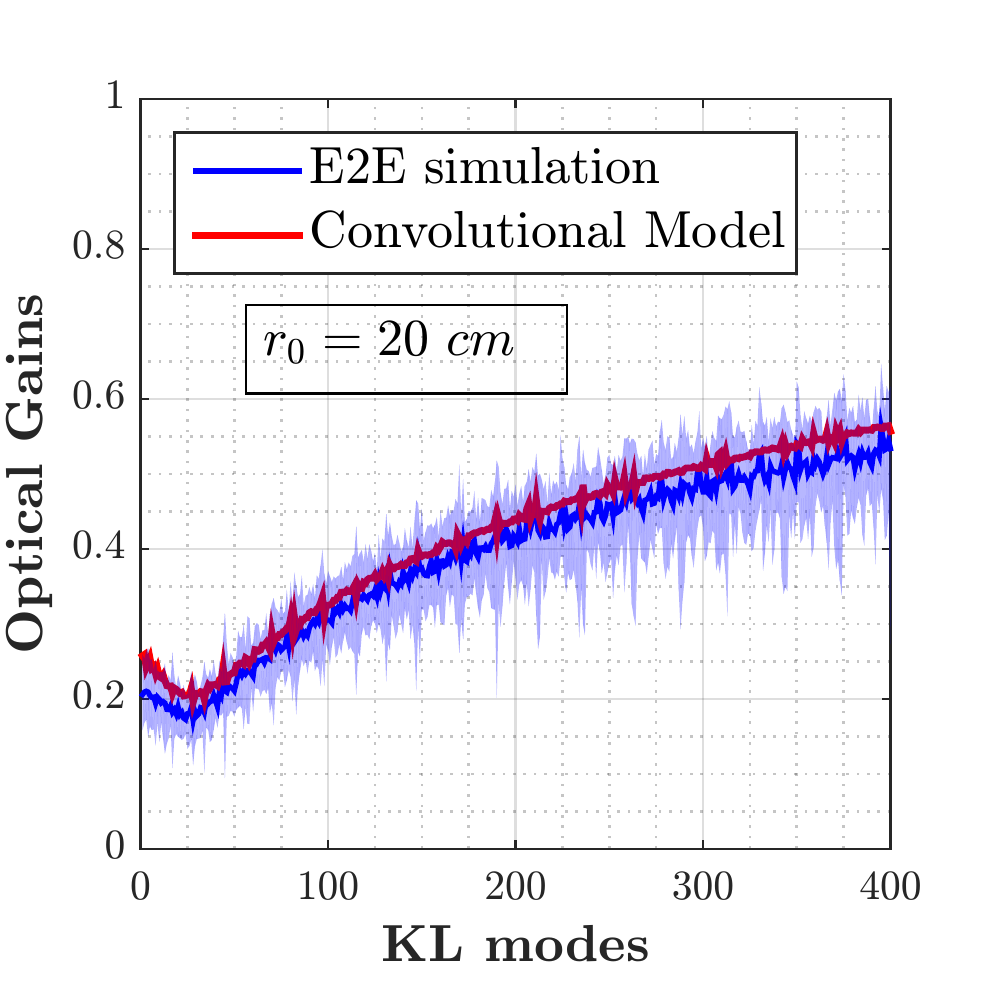}\\
    \includegraphics[scale=0.42]{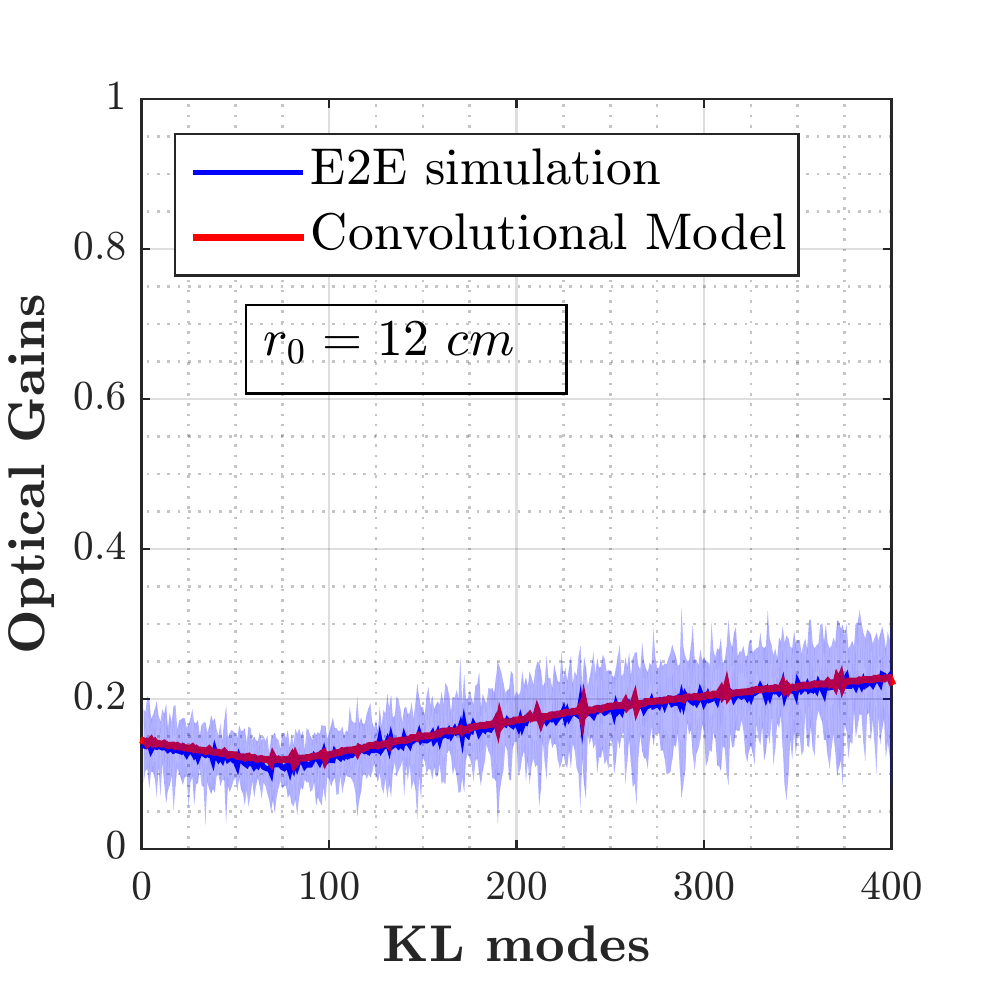}
    \includegraphics[scale=0.42]{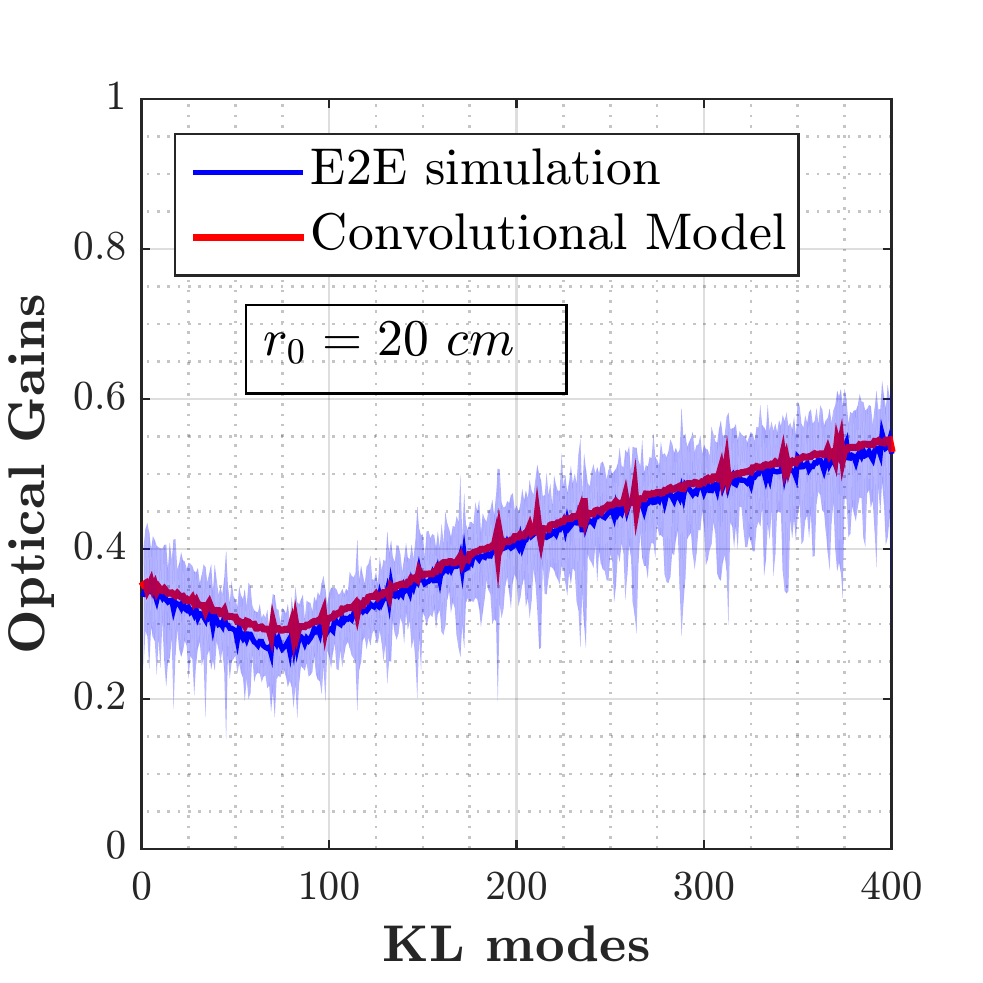}
    \caption{OG computed on full turbulence screens for different $r_{0}$. The convolutional model fits well with the OG computed by E2E simulations. The shaded area represents the maximum and minimum values found for the OG for 20 phase realisations.\textbf{Top}: $r_{mod} = 3\lambda/D$. \textbf{Bottom}: $r_{mod} = 5\lambda/D$.}
    \label{fig:fullTurbulence}
\end{figure}

\begin{figure}[!h]
    \centering
    \includegraphics[scale=0.42]{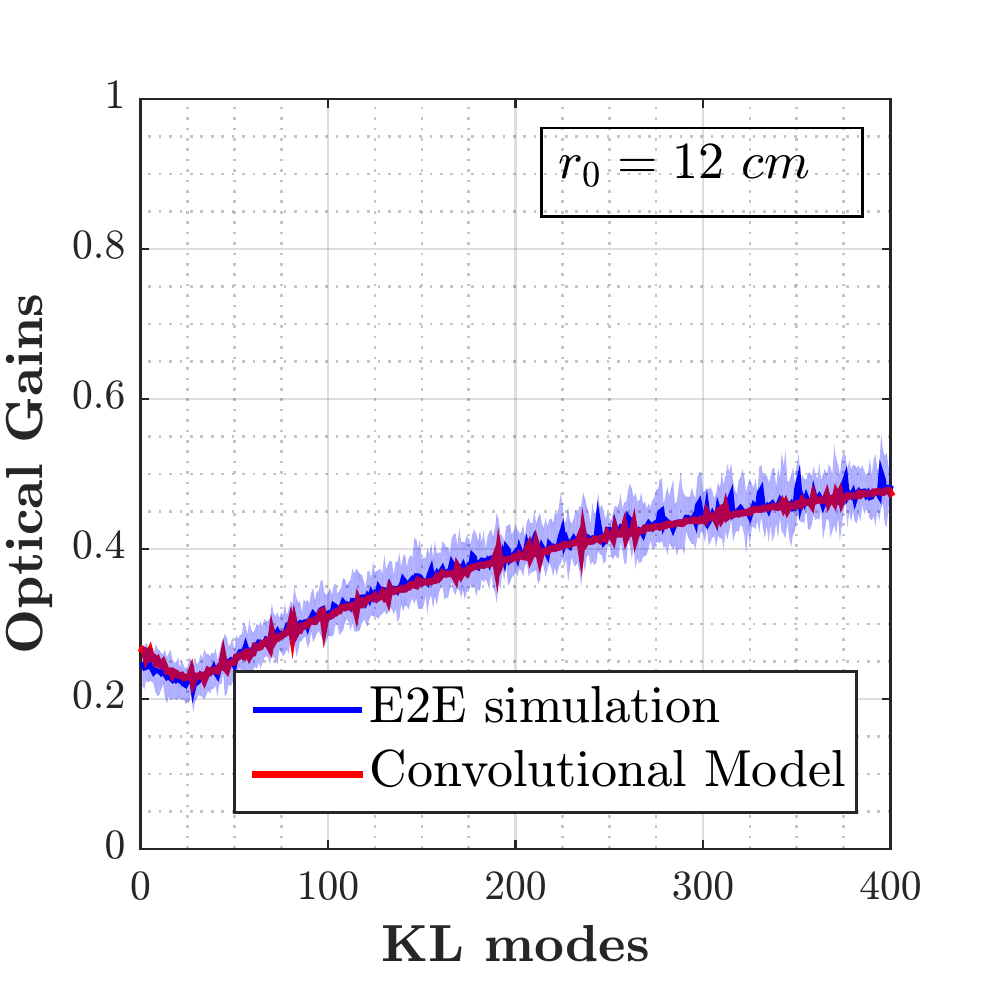}
    \includegraphics[scale=0.42]{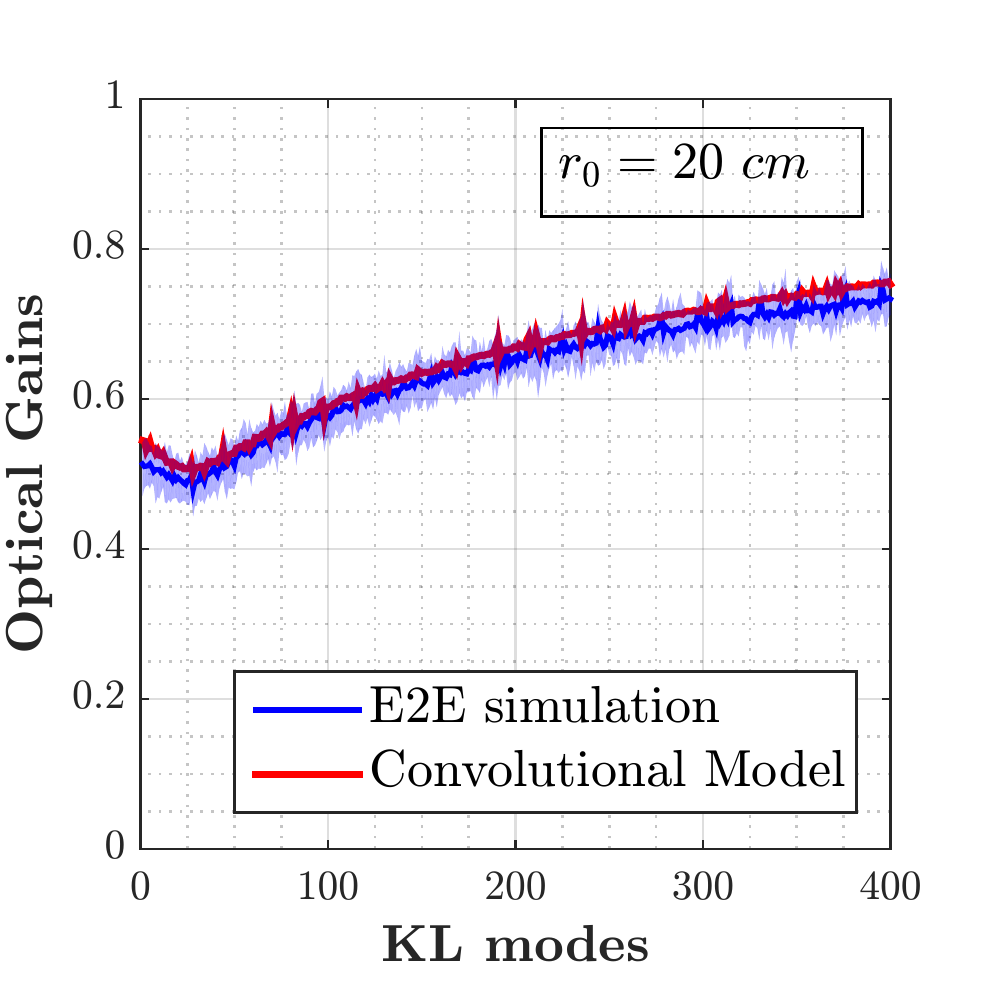}\\
    \includegraphics[scale=0.42]{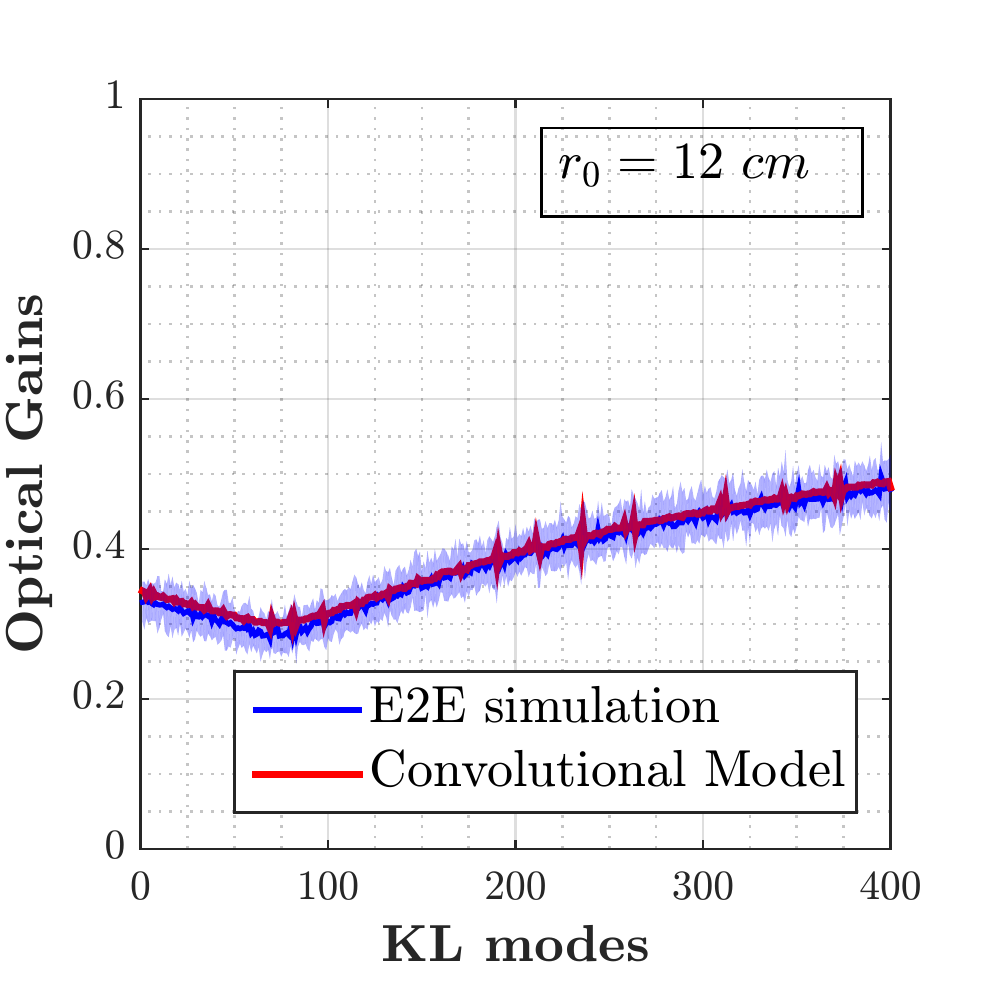}
    \includegraphics[scale=0.42]{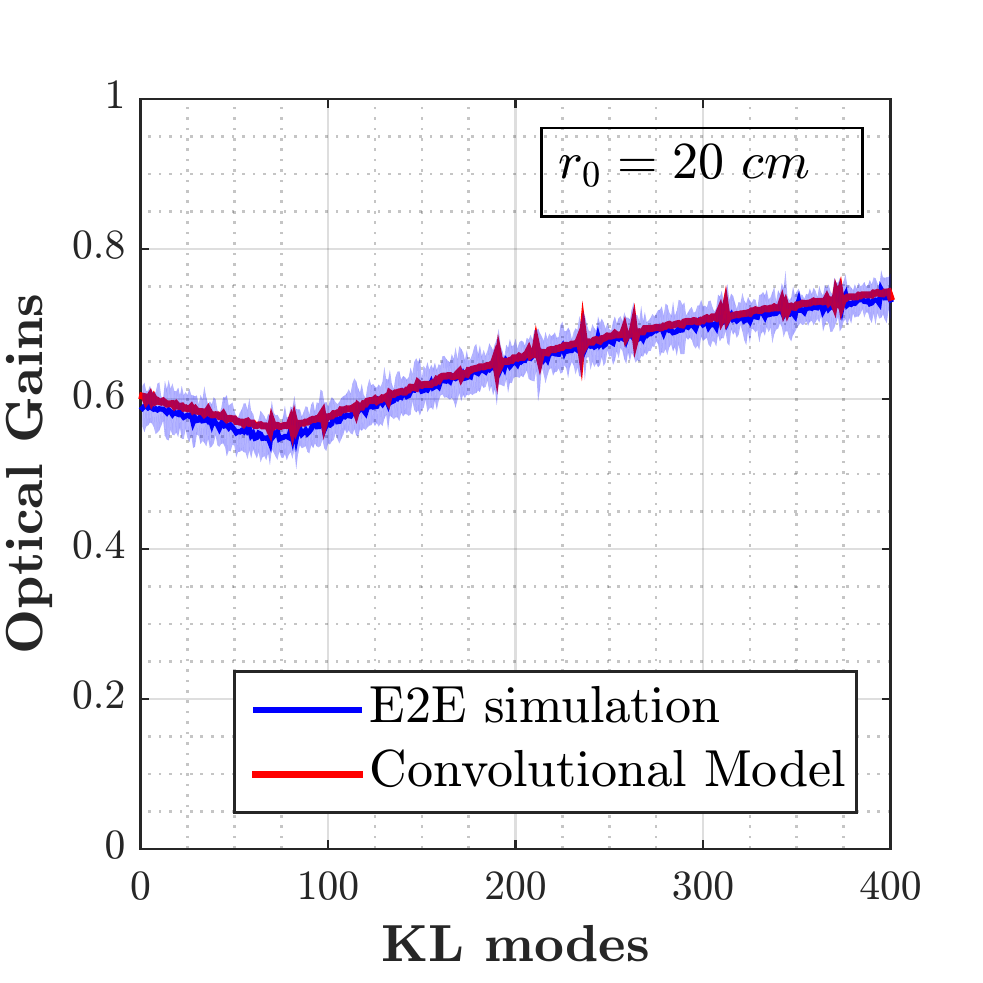}
    \caption{OG for closed loop residual phases. Number of actuators in the pupil: 20. \textbf{Top}: $r_{mod} = 3\lambda/D$. \textbf{Bottom}: $r_{mod}= 5\lambda/D$.}
    \label{fig:residualOG}
\end{figure}

\noindent\textbf{End-to-End simulations -} We proceed in the following way: given a PSD, we generate 20 decorrelated phases. We then compute the interaction matrices $IM_{\text{onSky}}$ around each of these phases (using a push-pull method) and we use the equation \ref{eq:OGmatrix} to compute the OG. The averaged values for each different PSD chosen are presented figure \ref{fig:fullTurbulence} and figure \ref{fig:residualOG} (the shaded areas represent the maximum and minimum values found for the OG for 20 phase realisations).

\noindent\textbf{Convolutional Model -} We use exploit the same PSD used for the End-to-End simulations to compute the $\textbf{IR}_{\textbf{onSky}}$ equation \ref{eq:IRsensing} and we retrieve the OG thanks to equation \ref{eq:Gconv}.\\

We can define two main PSD configurations around which we can compute the OG:

\begin{itemize}
    \item[$\bullet$]  The full turbulence OG: in that case, the PyWFS works in open-loop and wavefront sensing is done on a seeing-limited EM field at the apex of the pyramid. In the vast majority of systems, this is the case for the first loop iteration and before the loop is closed. After a few closed-loop iterations, the EM field seen by the pyramid is no longer seeing-limited because we are in closed-loop operation. We then fall in the second configuration described below. Tracking and compensating the OG in the full turbulence can be interesting when the system has convergence issues under strong turbulence or when we want to close the loop using low modulation radii. The results of the comparison for this configuration are given figure \ref{fig:fullTurbulence}: we note the strong agreement between the convolutional model and the End-to-End simulations.
    \item[$\bullet$]  The residual phases OG: the adaptive optics loop is closed and the OG are introduced by the imperfect wavefront correction. This case is the most interesting one because it can allow us to enhance the closed-loop performance. For this setting, the results are given figure \ref{fig:residualOG}: we still have a good match between our model and the End-to-End simulations.
\end{itemize}

By testing our model for different system configurations (different modulation radii, different $r_{0}$, open or closed loop residual phases) we have demonstrated that the convolutional model can be used to predict the OG with sufficient accuracy to remain in their statistical variability range. It therefore provides a fast and agile way for tracking OG, provided knowledge of the residual PSD. In the next section, we hence focus on how to get this information in a practical way.

\subsection{How to get the residual PSD?}

\begin{figure}[!h]
    \centering
    \includegraphics[scale=0.4]{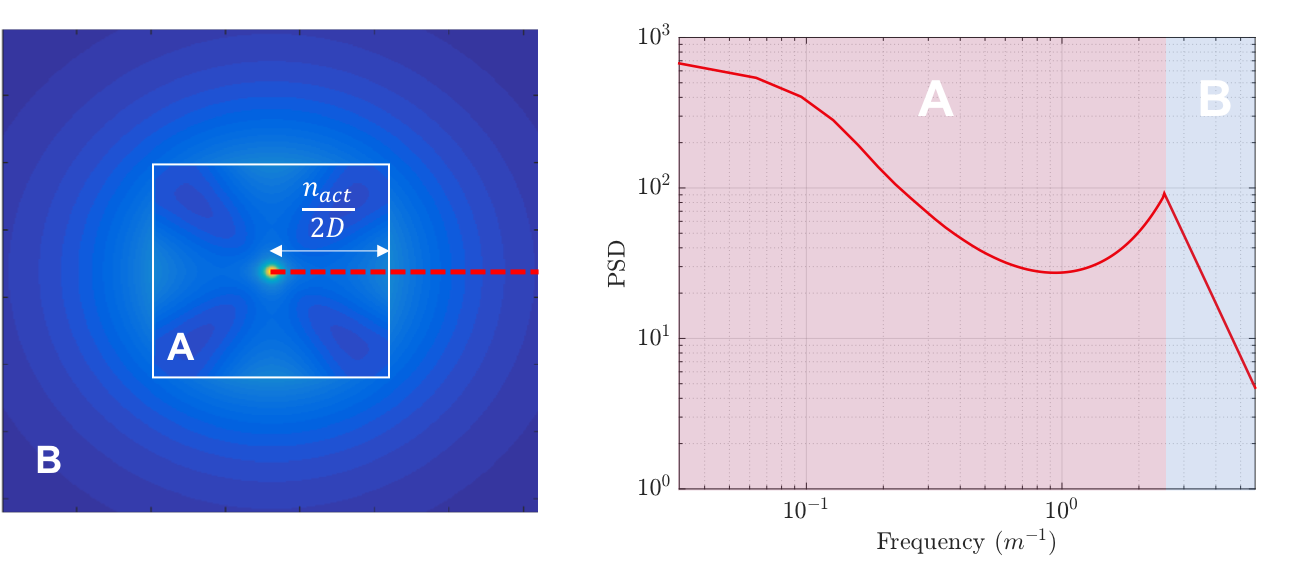}
    \caption{\textbf{Left:}Example of a residual PSD for a $40\times 40$ actuators system with a Cartesian geometry. In the frequency space, the correction zone is the area labelled A: it is a square with each side of $n_{act}=40/D$ in $m^{-1}$. The B area represents the space of uncorrected frequencies.\textbf{Right:}Radial cut of the PSD in log scale.}
    \label{fig:PSD}
\end{figure}

We propose here to obtain the residual PSD from the telemetry data. It is a non-invasive method that is already deeply investigated in the PSF reconstruction field \citep{Beltramo}. We remind the reader that the residual phase PSD can be split into two parts \citep{rigaut98}: the corrected frequencies (area A figure \ref{fig:PSD}) and the uncorrected frequencies (area B figure \ref{fig:PSD}). These two areas are separated by the Deformable Mirror (DM) cut-off frequency, which depends on the position and number of actuators. The PSD estimation process works in two steps:

\begin{itemize}
    \item[$\bullet$]  By recording the integrated commands sent to the DM, we can assess the shape of the turbulence. In other words, we are able to estimate the Fried parameter $r_{0}$ and therefore have an estimation of the shape of the PSD outside the correction zone. The estimation of $r_{0}$ thanks to telemetry data is usually not perfectly accurate and the Fried parameter is often overestimated. However, it has been shown that an AO system can be well characterized in order to correct for this offset \citep{fetick}.
    \item[$\bullet$]  Recording the residual commands will provide information on the residual PSD inside the correction area. This method is not ideal, because all the commands sent to the DM are already tainted by the OG problem. It is possible to overcome this issue by using models describing the analytical PSD inside the correction area, provided a simple set of parameters describing the system \citep{rigaut98,carlos}.
\end{itemize}

The next step is to understand what level of accuracy is required when computing the residual phase PSD using telemetry data combined with an analytical model of our system. Using the convolutional model, we propose a brief study to analyse the contribution of the different parts of residual PSD on the \textbf{OG morphology}. As we mentioned earlier, we can split the contribution of the residual phases into two parts: the fitting PSD and the PSD inside the correction zone. It is therefore interesting to study the OG for each of these contributors. \\

\begin{figure}[!h]
\centering
\includegraphics[scale=0.42]{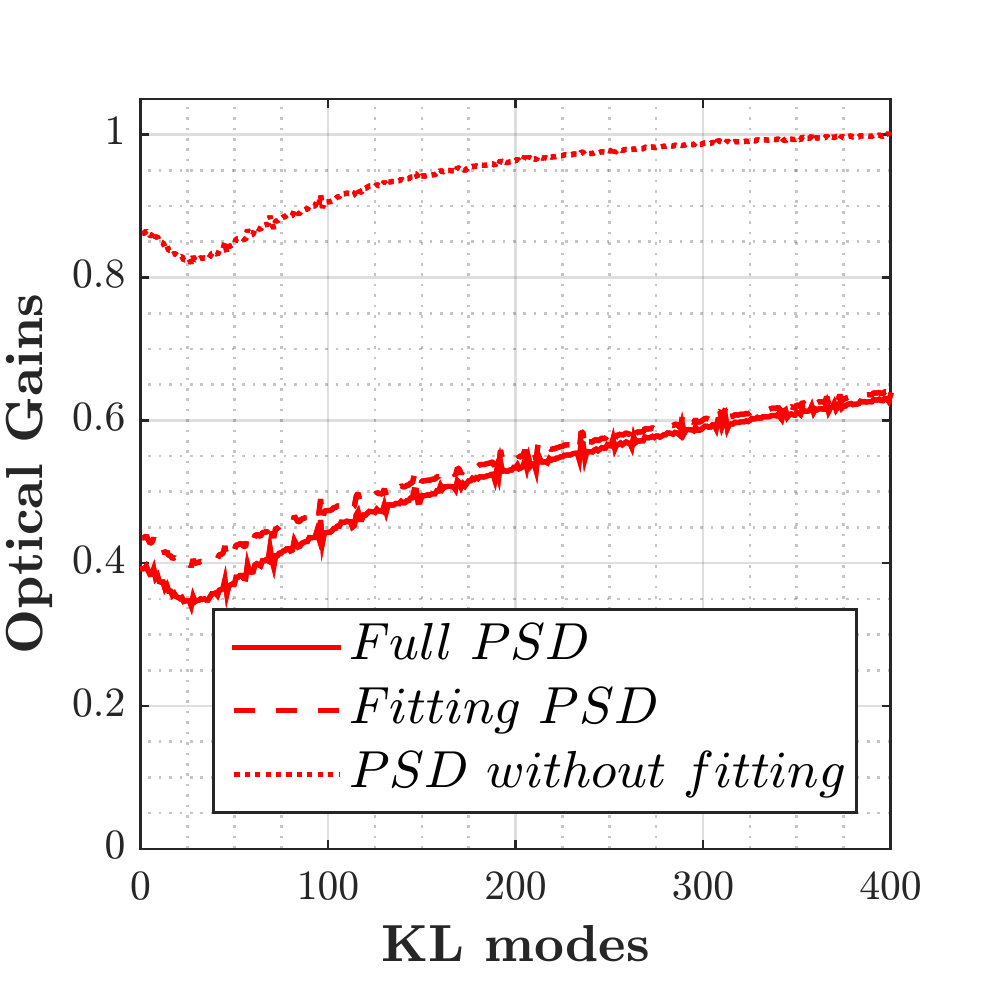}
\includegraphics[scale=0.42]{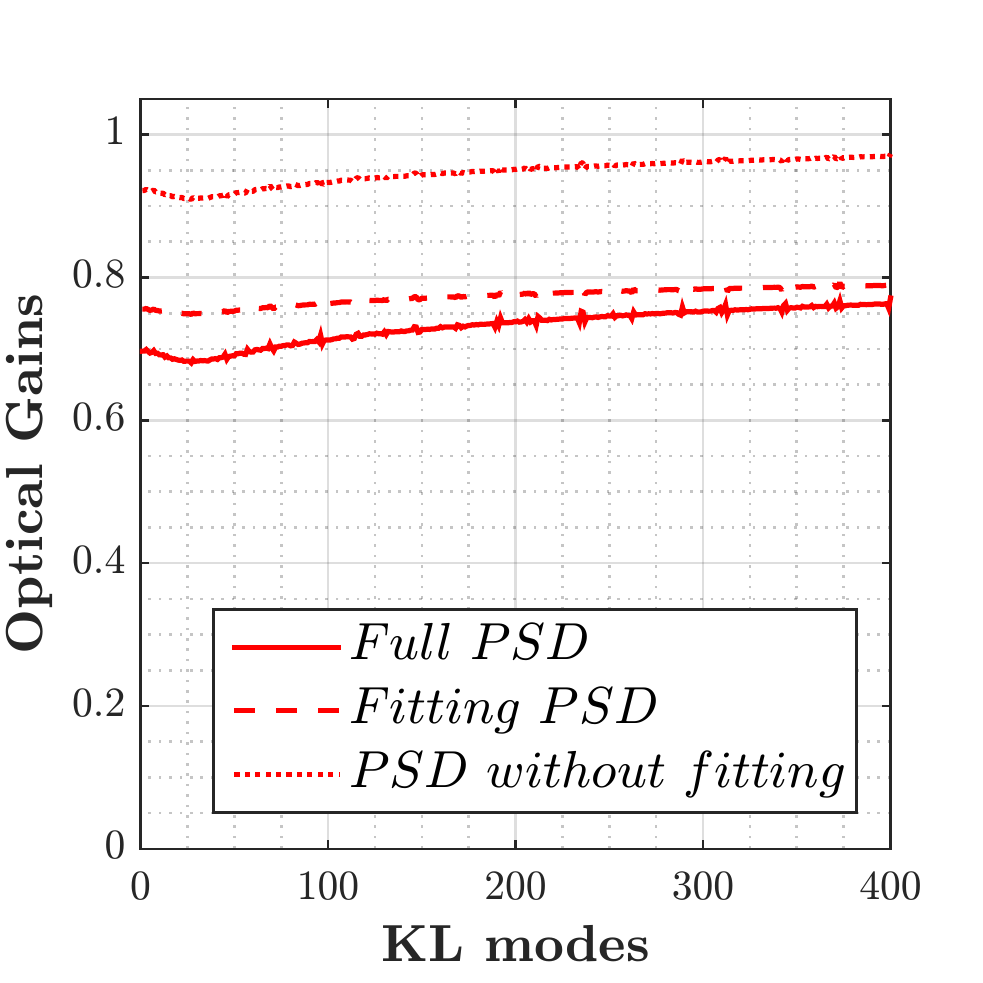}
\caption{Contribution of the corrected and uncorrected part of a closed loop PSD ($r_{0} = 15\ cm$) to the OG. \textbf{Left}: For 20 actuators in the pupil - NAOS configuration. \textbf{Right}: For 40 actuators in the pupil - SPHERE configuration.}
\label{fig:fittingPSD}
\end{figure}

\begin{figure}[!h]
\centering
\includegraphics[scale=0.6]{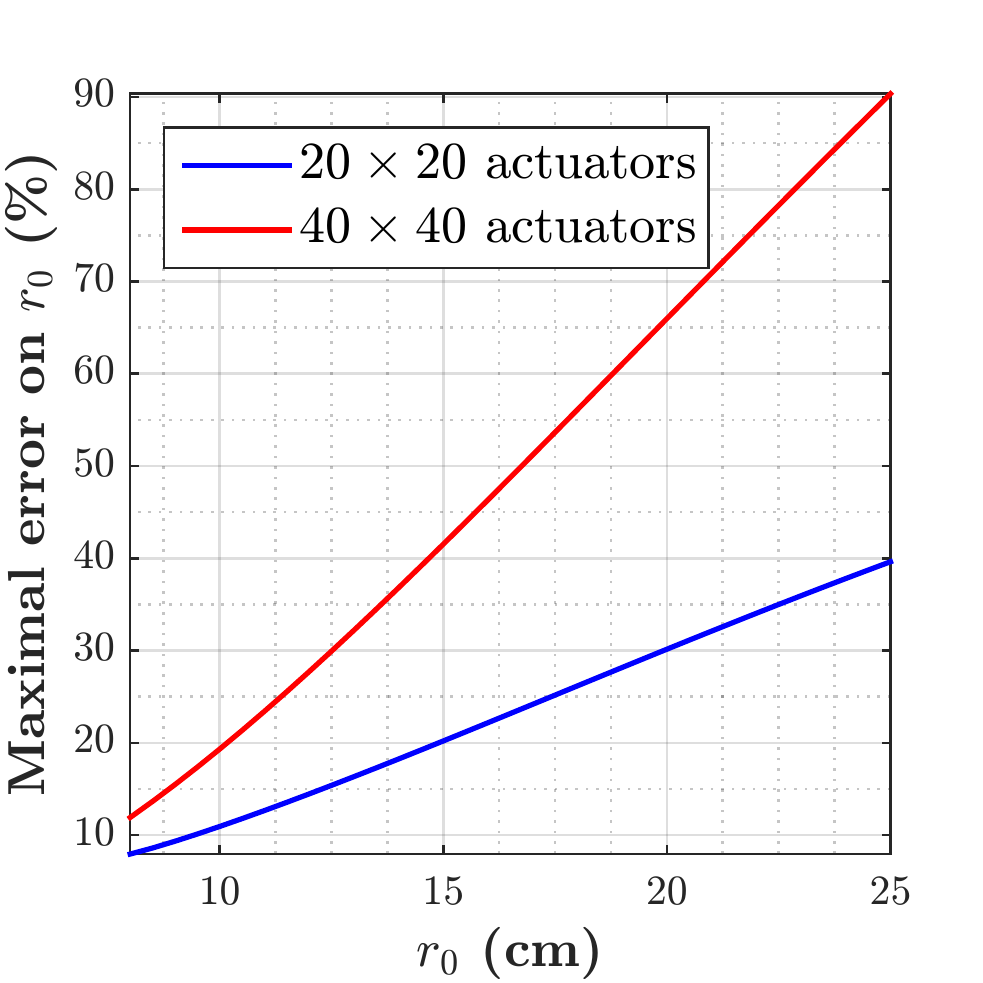}
\caption{Maximum acceptable error (in percent) on the estimation of $r_{0}$ to ensure an error on the computed OG under $\pm 10\%$ for two different system configurations.}
\label{fig:errorOnr0}
\end{figure}

For that purpose we choose two system configurations: a 8m telescope given a $r_{0} =15\ cm$ with either 20 actuators (NAOS-like configuration on the VLT) or 40 actuators within the same pupil diameter (SPHERE-like configuration on the VLT: \cite{sphere}). We use typical residual PSD of these systems to compute the OG thanks to the convolutional model. Figure~\ref{fig:fittingPSD}, we show results when the OG are computed for the full PSD, for the fitting part of the PSD and for the PSD inside the correction area only. For these chosen configurations, it is clear that OG gains are dominated by the energy which lies in the fitting PSD, even for the high-contrast configuration ($40\times 40$ actuators in the pupil) where the residual energy is equally distributed between the corrected and the uncorrected zone (figure \ref{fig:fittingPSD}). Therefore the previous statement tends to be often verified (all the more because we are considering residual phases at the wavefront sensing wavelengths). Yet, it is clear that for a very noisy AO system, the fitting error could be overcome by the error inside the correction zone: in that case, the error on the OG computation will be constrain by the estimation of the residual PSD inside the correction zone. Nevertheless, we can conclude that in the vast majority of the observations and for present and future AO systems (the E-ELT will also be in a fitting error limited AO configuration), the OG morphology is mainly constrained by the Fried parameter $r_{0}$, and that the knowledge of this parameter only would be enough to derive a sufficiently accurate model of the OG. Thus, estimating $r_{0}$ during closed-loop operation is a crucial step for PyWFS OG tracking. In order to assess the accuracy on $r_{0}$ that needs to reached, we probe what impact an error in the estimation of $r_{0}$ has on the computation of OG in figure \ref{fig:errorOnr0}. In this plot, and for both configurations studied, we present the maximal acceptable error on the estimation of $r_{0}$ to maintain an error on computed OG below $\pm 10\%$. In order to retrieve OG with an error below $\pm 10\%$, we see that we need to be more accurate for bad seeing conditions and for AO systems with less DM actuators in the pupil. Overall, the values presented in this figure show that we do not need an incredibly high precision on the Fried parameter to accurately compute the OG using the presented method.

\section{Applying the convolutional model to NCPA correction}
\label{NCPA}

\begin{figure}[!h]
    \centering
    \includegraphics[scale=0.6]{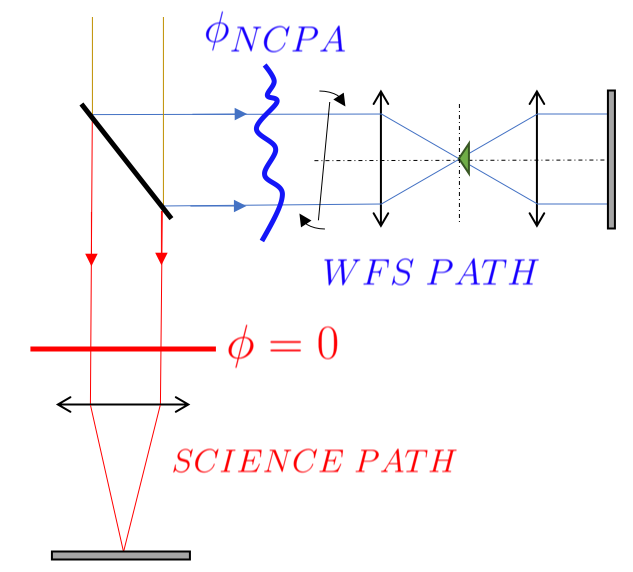}
    \caption{In order to obtain a flat wavefront on the science path, the zero-point wavefront on the WFS needs to correspond to the NCPA.}
    \label{fig:ncpa}
\end{figure}

The aim of this section is to demonstrate the importance of estimating OGs for the correct control of AO system by focusing on the specific issue of NCPA correction. NCPA appear in AO systems when the aberrations between the WFS path and science path are different. In that case, if nothing is done, the AO loop converges towards a flat wavefront on the WFS and the NCPA remain uncorrected on the science camera. This effect can be mitigated by using a non-null wavefront reference target on the WFS, corresponding to the NCPA (figure \ref{fig:ncpa}). To do so, we propose to proceed with the following three calibration steps:

\begin{itemize}
    \item[1 - ] Determination of the NCPA wavefront (using techniques such as phase diversity for instance \citep{blanc2003}).
    \item[2 - ] Computation of the Interaction Matrix $IM_{calib}$ around the NCPA wavefront. Because NCPA are not a zero-mean stationary wavefront, they cannot be described by the convolutional model through a structure function as is stated equation \ref{eq:IRsensing}. Furthermore, the diagonal approximation (section 2.3) is not necessarily verified in the case of NCPA. Hence, it is better to calibrate the WFS as close as possible to its working point: the NCPA wavefront. 
    \item[3 - ] Computation of the WFS response to the NCPA wavefront $I_{calib}(\phi_{NCPA})$.
\end{itemize}

Subsequently, the reference WFS intensities correspond to the NCPA. Given a residual phase $\phi_{res}$, the signal to be reconstructed is then $I_{calib}(\phi_{res}) - I_{calib}(\phi_{NCPA})$. However, using this strategy for PyWFS in presence of residual phase OG is unfortunately problematic and can lead to critical loop instabilities.

\subsection{The NCPA catastrophe}

\begin{figure}[!h]
    \centering
    \includegraphics[scale=0.5]{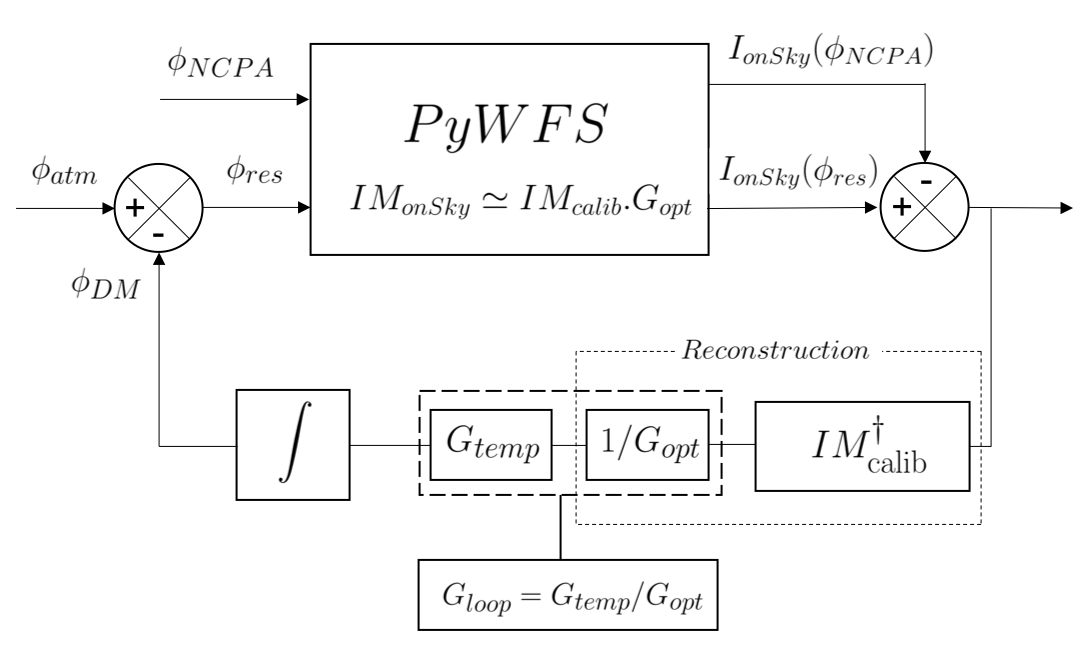}
    \caption{Schematic view of the AO closed-loop in presence of compensated NCPA. The feed-back loop can be otpimized by computing the quantity $G_{loop}$ without disentangling $G_{temp}$ from $G_{opt}$. However, to properly compensate for the NCPA in the foward loop, the value $G_{opt}$ is needed.}
    \label{fig:ncpaAutomatic}
\end{figure}

For a WFS working around its reference position, the signal to be reconstructed is $I_{calib}(\phi_{res}) - I_{calib}(\phi_{NCPA})$. In the case of a classical integral controller, the commands sent to the DM at each frame $t$ is:

\begin{equation}
     c(t) = c(t-1) - G_{temp}.IM_{calib}^{\dag}.[I_{calib}(\phi_{res}(t)) - I_{calib}(\phi_{NCPA})]
     \label{eq:DMcommands}
\end{equation}

where $G_{temp}$ is a diagonal matrix, ideally constituted of the optimized temporal modal gains of the loop. This equation works for a perfectly linear WFS. As it was previously presented in this paper, the PyWFS exhibits OG. Therefore, the on-sky PyWFS measurements are: 

\begin{equation}
     PyWFS:\phi \rightarrow I_{onSky}(\phi)
\end{equation}

thus, the equation \ref{eq:DMcommands} becomes:

\begin{equation}
     c(t) = c(t-1) - G_{temp}.IM_{onSky}^{\dag}.[I_{onSky}(\phi_{res}(t)) - I_{onSky}(\phi_{NCPA})]
\end{equation}

which gives in the OG diagonal approximation:

\begin{equation}
     c(t) = c(t-1) - G_{temp}.\frac{IM_{calib}^{\dag}}{G_{opt}}.[I_{onSky}(\phi_{res}(t)) - I_{onSky}(\phi_{NCPA})]
\end{equation}

where $G_{loop}=G_{temp}/G_{opt}$ is therefore a diagonal matrix used to apply different gains on each of the controlled modes. As we mentioned in the introduction, very efficient methods are avaiable that can optimize this matrix (\cite{close}), but without differentiating between $G_{temp}$ and $G_{opt}$. However, it is important to notice in this equation that the intensity to be removed is $I_{onSky}(\phi_{NCPA})$ and not $I_{calib}(\phi_{NCPA})$ (see figure \ref{fig:ncpaAutomatic}). These two quantities are linked through the following equation:

\begin{equation}
\begin{split}
    I_{onSky}(\phi_{NCPA}) = IM_{onSky}.\phi_{NCPA}\\
    I_{onSky}(\phi_{NCPA}) = IM_{calib}.G_{opt}.IM_{calib}^{\dag}.I_{calib}(\phi_{NCPA})
\end{split}
\label{eq:Incpa}
\end{equation}

We see from this equation that we need to get $G_{opt}$ to be able to properly compensate for the NCPA.\\ 

\begin{figure}[!h]
    \centering
    \includegraphics[scale=0.4]{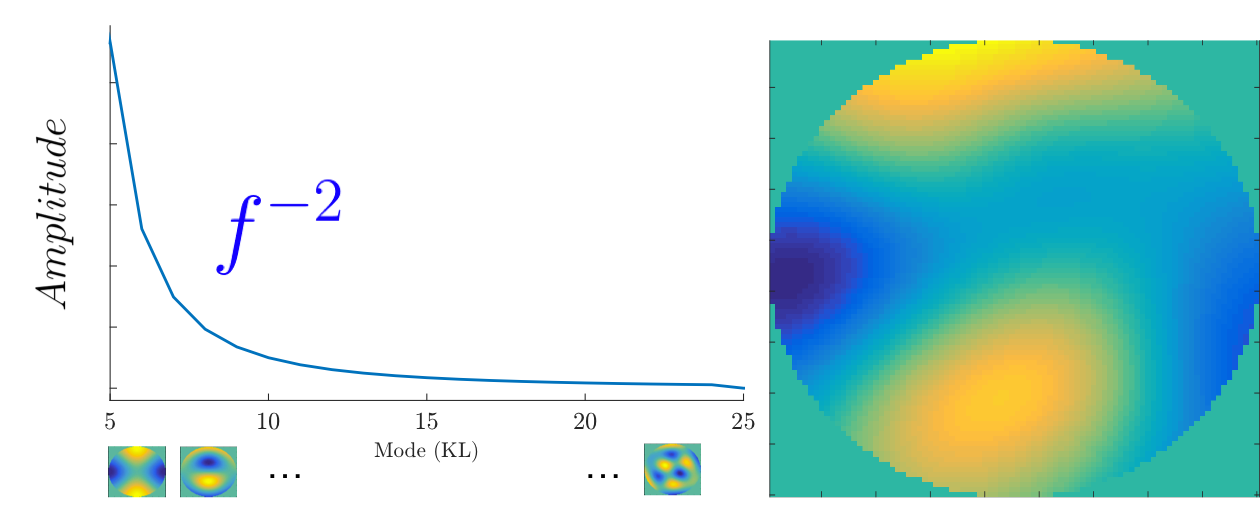}
    \caption{The NCPA phase chosen for our simulations is a linear combination of 20 low-order KL modes following a $f^{-2}$ law in rms amplitude.}
    \label{fig:ncpaPhase}
\end{figure}

We can wonder what would happens if we were just to use $I_{calib}(\phi_{NCPA})$ in equation \ref{eq:Incpa}. To answer this question, we performed End-to-End simulations with the same parameters as before, that is to say for a 8m telescope with 400 controlled KL modes and a fried parameter $r_{0} = 15\ cm$. The wavelength of the science path is chosen to be the H-band. Because NCPA are usually composed low-order modes, we choose the following arbitrary distribution for the NCPA: a combination of the modes KL5 to KL25, following a $f^{-2}$ law (see figure \ref{fig:ncpaPhase}).

We run several close loop simulations while increasing the NCPA amplitudes, and we record the Strehl Ratio over 16 seconds of close loop integration. The results are given figure \ref{fig:ncpaCatastrophe}. The dashed line shows the impact of increasing NCPA in the case where we do not try to compensate for them. When we try to compensate the NCPA by applying reference intensities on the PyWFS without compensating for the OG, we observe a degradation of performance. This is not surprising: when subtracting the NCPA reference intensities to the PyWFS measurements, the mode $\phi_{NCPA,i}$ will be reconstructed:

\begin{equation}
\begin{split}
    \widetilde{\phi}_{NCPA,i} = IM_{\text{onSky}}^{\dag}.I_{calib}(\phi_{NCPA,i})= \frac{\phi_{NCPA,i}}{g_{opt}(\phi_{NCPA,i})}
\end{split}
\label{eq:ncpaPush}
\end{equation}

where $g_{opt}(\phi_{NCPA,i})<1$ is the OG associated with the mode $\phi_{NCPA,i}$, and so we have:

\begin{equation}
\begin{split}
    \widetilde{\phi}_{NCPA,i}>\phi_{NCPA,i}
\end{split}
\label{eq:ncpaPush}
\end{equation}

\begin{figure}[!h]
    \centering
    \includegraphics[scale=0.45]{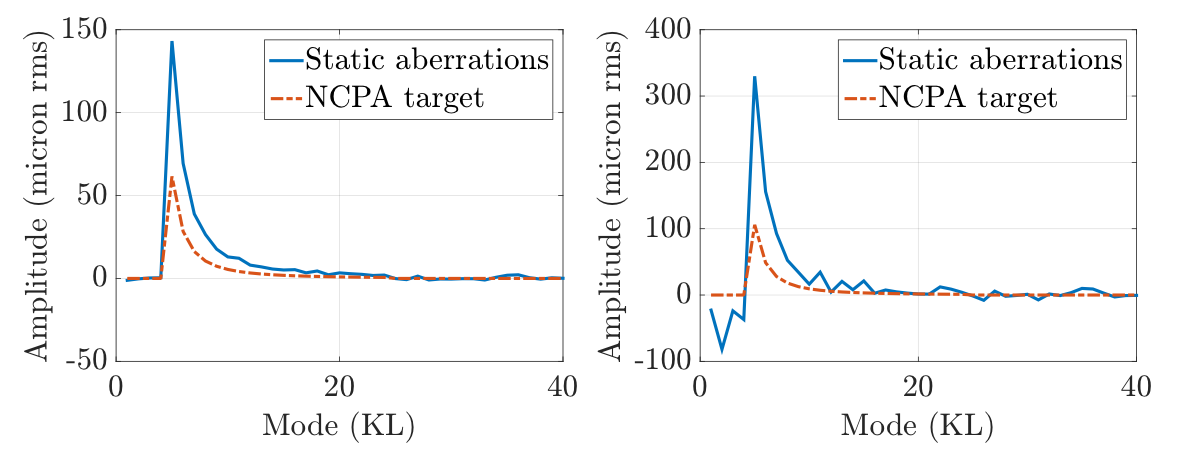}
    \caption{Static aberrations in the AO loop in the case of no OG compensation on the NCPA reference intensities. \textbf{Left:} for NCPA of 70 nm rms.\textbf{Right:} for NCPA of 120 nm rms.}
    \label{fig:ncpaStatic}
\end{figure}

It emphasises the fact that if we do not compensate for OG, the reference intensities will produce an excess of NCPA proportional to the OG in the loop. This effect is shown figure \ref{fig:ncpaStatic} for two cases highlighted by red circles figure \ref{fig:ncpaCatastrophe}.

When the amplitude of the NCPA is sufficiently high, producing too much NCPA will create additional OG that will add to the ones already generated by the residual phases. By lowering the OG, it will lead to an increase of NCPA correction according to equation  \ref{eq:ncpaPush}. The increase in NCPA correction will significantly change the OG and lead to ever higher NCPA correction levels. This will make the loop diverge: it is what we labelled the NCPA catastrophe. We can clearly see this effect in our simulations starting from $130\ nm\ rms$ and above of NCPA (figure \ref{fig:ncpaCatastrophe}).

\begin{figure}[!h]
    \centering
    \includegraphics[scale=0.75]{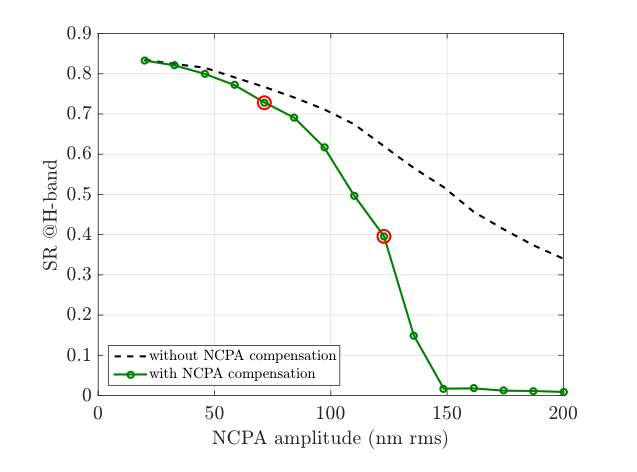}
    \caption{Strehl Ratio for increasing NCPA amplitude, in the case of no NCPA compensation and of no OG compensation on the NCPA reference intensities. We notice that for too high NCPA amplitude when the reference intensities are not updated with OG, the loop diverges: this is what we have labelled the NCPA catastrophe. The static aberrations of the two configurations marked by a red circle are plotted figure \ref{fig:ncpaStatic}.}
    \label{fig:ncpaCatastrophe}
\end{figure}

 \subsection{NCPA compensation using the convolutional model}
 
If we suppose the residual phase PSD a known quantity, we can use the convolutional model to compute the OG and to update the reference intensities according to equation \ref{eq:Incpa}. By doing so, we obtain the upper curve figure \ref{fig:ncpaCatastrophe}. Performance is significantly improved, but there is still a built up of static aberrations during the closed-loop operation, preventing the system to maintain its maximum Strehl ratio irrespective of NCPA amplitude (which would correspond to a flat curve figure \ref{fig:ncpaCatastrophe}). This can be explained by two phenomena:

\begin{itemize}
    \item[$\bullet$]  The way the OG has been defined corresponds to an average state of the system (equation \ref{eq:average}). At each frame, the current OG can be higher or lower than the averaged value, introducing an error on the NCPA reference intensities. The ideal strategy would be to have the means to estimate the OG at each frame.
    \item[$\bullet$]  The convolutional model characterises the offset between PyWFS measurements when the calibrating around a null-phase and when in presence of residual phases. But it does not take in account the presence of NCPA in the shape of the computed OG. Therefore, with higher NCPA amplitudes the error on the OG computed with the convolutional model is increased. This explains why performance is decreased with increasing NCPA amplitudes in figure \ref{fig:ncpaCatastrophe}. Further analysis of this problem is beyond the scope of this paper, but we are currently working on a solution that requires further analytical developments on the convolutional model.
\end{itemize}

\begin{figure}[!h]
    \centering
    \includegraphics[scale=0.75]{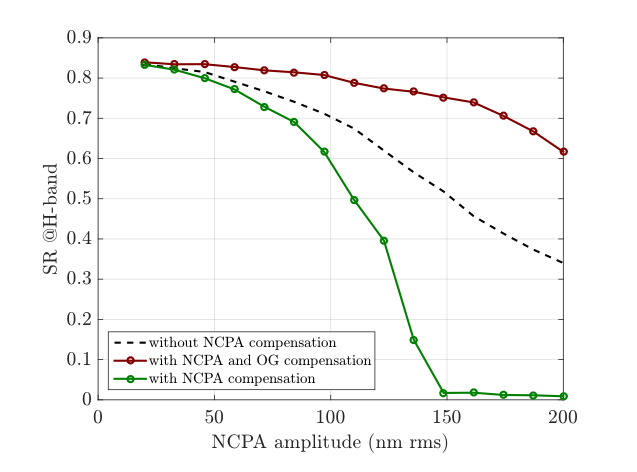}
    \caption{Strehl Ratio for increasing NCPA amplitude in the case of OG compensation on the NCPA reference intensities, compared to the previous cases presented before. The performance is increased and the NCPA catastrophe avoided. Nonetheless, a noticeable impact on performance is visible as the NCPA amplitude is increased.}
    \label{fig:ncpaCatastrophe}
\end{figure}

This section highlights the importance of estimating the PyWFS OG for NCPA compensation in close loop operation. The OG estimation based on the convolutional model has proven to be efficient for typical NCPA amplitudes (below 100 nm rms) encountered in AO systems. However, handling stronger NCPA amplitudes will require further analytical developments in order to take into account the modification of OG by the NCPA themselves.

\section{Conclusions}

The work presented in this paper offers a new method for computing the Pyramid WFS Optical Gains. Our approach relies on a physical description of the wavefront sensor through a convolutional model, which allows to analytically compute the impact of residual phases on PyWFS measurements. We have demonstrated the accuracy of this method by comparing results to End-to-End simulations for different system configurations.

The presented method requires knowledge of the residual phase statistical characteristics to compute the OG. We have presented a practical implementation to estimate residual phase statistics using AO telemetry data, in a similar to what is done for PSF reconstruction. We have showed that the most important aspect is the knowledge of the turbulence strength through the Fried parameter $r_{0}$. We also demonstrated that from this $r_{0}$ parameter alone, a good approximation of the OG could be achieved. In other words, any AO system using a pyramid WFS and capable of providing an online estimate of $r_{0}$ could benefit from estimating the OG using the method we presented in this paper.

Finally, we demonstrated that the OG play a crucial part when trying to compensate NCPA with a PyWFS. To avoid what we have labelled the NCPA catastrophe, proper handling of the OG is mandatory. We have proposed a way to mitigate the impact of OG on NCPA by computing them using the method we presented in this paper.
In fact, this work can also be applied to any type of wavefront sensor based on Fourier-filtering, and provides a new insight into the understanding of OG in Fourier Filtering WFS and how to manage them.\\

\noindent \textbf{Acknowledgments}\\
\noindent 
This document has been prepared as part of the activities of OPTICON H2020 (2017-2020) Work Package 1 (Calibration and test tools for adaptive-optics assisted E-ELT instruments). OPTICON is supported by the Horizon 2020 Framework Programme of the European Commission's (Grant number 730890).  This work was supported by the Action Spécifique Haute Résolution Angulaire (ASHRA) of CNRS/INSU co-funded by CNES. This work also benefited from the support of the WOLF project ANR-18-CE31-0018 of the French National Research Agency (ANR).

\bigskip

\bibliographystyle{aa} 
\bibliography{OGpaper}

\begin{thebibliography}{24}
\expandafter\ifx\csname natexlab\endcsname\relax\def\natexlab#1{#1}\fi

\bibitem[{Beltramo-Martin {et~al.}(2019)Beltramo-Martin, Correia, Ragland,
  Jolissaint, Neichel, Fusco, \& Wizinowich}]{Beltramo}
Beltramo-Martin, O., Correia, C.~M., Ragland, S., {et~al.} 2019, Monthly
  Notices of the Royal Astronomical Society, 487, 5450–5462

\bibitem[{{Beuzit} {et~al.}(2019){Beuzit}, {Vigan}, {Mouillet}, {Dohlen},
  {Gratton}, {Boccaletti}, {Sauvage}, {Schmid}, {Langlois}, {Petit},
  {Baruffolo}, {Feldt}, {Milli}, {Wahhaj}, {Abe}, {Anselmi}, {Antichi},
  {Barette}, {Baudrand}, {Baudoz}, {Bazzon}, {Bernardi}, {Blanchard}, {Brast},
  {Bruno}, {Buey}, {Carbillet}, {Carle}, {Cascone}, {Chapron}, {Charton},
  {Chauvin}, {Claudi}, {Costille}, {De Caprio}, {de Boer}, {Delboulb{\'e}},
  {Desidera}, {Dominik}, {Downing}, {Dupuis}, {Fabron}, {Fantinel}, {Farisato},
  {Feautrier}, {Fedrigo}, {Fusco}, {Gigan}, {Ginski}, {Girard}, {Giro},
  {Gisler}, {Gluck}, {Gry}, {Henning}, {Hubin}, {Hugot}, {Incorvaia}, {Jaquet},
  {Kasper}, {Lagadec}, {Lagrange}, {Le Coroller}, {Le Mignant}, {Le Ruyet},
  {Lessio}, {Lizon}, {Llored}, {Lundin}, {Madec}, {Magnard}, {Marteaud},
  {Martinez}, {Maurel}, {M{\'e}nard}, {Mesa}, {M{\"o}ller-Nilsson}, {Moulin},
  {Moutou}, {Orign{\'e}}, {Parisot}, {Pavlov}, {Perret}, {Pragt}, {Puget},
  {Rabou}, {Ramos}, {Reess}, {Rigal}, {Rochat}, {Roelfsema}, {Rousset}, {Roux},
  {Saisse}, {Salasnich}, {Santambrogio}, {Scuderi}, {Segransan}, {Sevin},
  {Siebenmorgen}, {Soenke}, {Stadler}, {Suarez}, {Tiph{\`e}ne}, {Turatto},
  {Udry}, {Vakili}, {Waters}, {Weber}, {Wildi}, {Zins}, \& {Zurlo}}]{sphere}
{Beuzit}, J.~L., {Vigan}, A., {Mouillet}, D., {et~al.} 2019, \aap, 631, A155

\bibitem[{Blanc {et~al.}(2003)Blanc, Fusco, Hartung, Mugnier, \&
  Rousset}]{blanc2003}
Blanc, A., Fusco, T., Hartung, M., Mugnier, L., \& Rousset, G. 2003, Astronomy
  and Astrophysics, v.399, 373-383 (2003), 399

\bibitem[{Conan \& Correia(2014)}]{oomao}
Conan, R. \& Correia, C. 2014, in , 91486C

\bibitem[{Correia {et~al.}(2020)Correia, Fauvarque, Bond, Chambouleyron,
  Sauvage, \& Fusco}]{carlos}
Correia, C., Fauvarque, O., Bond, C., {et~al.} 2020, astro-ph.IM

\bibitem[{Davies {et~al.}(2018)Davies, Alves, Clénet, Lang-Bardl, Nicklas,
  Pott, Ragazzoni, Tolstoy, Amico, Anwand-Heerwart, Barboza, Barl, Baudoz,
  Bender, Bezawada, Bizenberger, Boland, Bonifacio, Borgo, Buey, Chapron,
  Chemla, Cohen, Czoske, Deo, Disseau, Dreizler, Dupuis, Fabricius, Falomo,
  Fedou, Schreiber, Garrel, Geis, Gemperlein, Gendron, Genzel, Gillessen,
  Glück, Grupp, Hartl, Häuser, Hess, Hofferbert, Hopp, Hörmann, Hubert,
  Huby, Huet, Hutterer, Ives, Janssen, Jellema, Kausch, Kerber, Kravcar, Ruyet,
  Leschinski, Mandla, Manhart, Massari, Mei, Merlin, Mohr, Monna, Muench,
  Mueller, Musters, Navarro, Neumann, Neumayer, Niebsch, Plattner, Przybilla,
  Rabien, Ramlau, Ramos, Ramsay, Rhode, Richter, Richter, Rix, Rodeghiero,
  Rohloff, Rosensteiner, Rousset, Schlichter, Schubert, Sevin, Stuik, Sturm,
  Thomas, Tromp, Kleijn, Vidal, Wagner, Wegner, Zeilinger, Ziegleder, Ziegler,
  \& Zins}]{micado}
Davies, R., Alves, J., Clénet, Y., {et~al.} 2018, The MICADO first light
  imager for the ELT: overview, operation, simulation

\bibitem[{{Deo} {et~al.}(2019){Deo}, {Gendron}, {Rousset}, {Vidal}, {Sevin},
  {Ferreira}, {Gratadour}, \& {Buey}}]{vdeo}
{Deo}, V., {Gendron}, {\'E}., {Rousset}, G., {et~al.} 2019, \aap, 629, A107

\bibitem[{Deo {et~al.}(2019)Deo, Rozel, Bertrou-Cantou, Ferreira, Vidal,
  Gratadour, A.~Sevin, Rousset, \& Gendron}]{close}
Deo, V., Rozel, M., Bertrou-Cantou, A., {et~al.} 2019

\bibitem[{Esposito {et~al.}(2015)Esposito, Pinna, Puglisi, Agapito, Veran, \&
  Herriot}]{espositoNCPA}
Esposito, S., Pinna, E., Puglisi, A., {et~al.} 2015in

\bibitem[{Fauvarque {et~al.}(2019)Fauvarque, Janin-Potiron, Correia,
  Br\^{u}l\'{e}, Neichel, Chambouleyron, Sauvage, \& Fusco}]{fauv}
Fauvarque, O., Janin-Potiron, P., Correia, C., {et~al.} 2019, J. Opt. Soc. Am.
  A, 36, 1241

\bibitem[{{F{\'e}tick} {et~al.}(2019){F{\'e}tick}, {Fusco}, {Neichel},
  {Mugnier}, {Beltramo-Martin}, {Bonnefois}, {Petit}, {Milli}, {Vernet},
  {Oberti}, \& {Bacon}}]{fetick}
{F{\'e}tick}, R.~J.~L., {Fusco}, T., {Neichel}, B., {et~al.} 2019, \aap, 628,
  A99

\bibitem[{Guyon(2005)}]{Guyon_2005}
Guyon, O. 2005, The Astrophysical Journal, 629, 592–614

\bibitem[{Hippler {et~al.}(2019)Hippler, Feldt, Bertram, Brandner, Cantalloube,
  Carlomagno, Absil, Obereder, Shatokhina, \& Stuik}]{metis}
Hippler, S., Feldt, M., Bertram, T., {et~al.} 2019, Experimental Astronomy

\bibitem[{Hutterer {et~al.}(2018)Hutterer, Ramlau, \& Shatokhina}]{vicky}
Hutterer, V., Ramlau, R., \& Shatokhina, I. 2018, Real-time Adaptive Optics
  with pyramid wavefront sensors: A theoretical analysis of the pyramid sensor
  model

\bibitem[{Korkiakoski {et~al.}(2007)Korkiakoski, Verinaud, Louarn, \&
  Conan}]{kkR}
Korkiakoski, V., Verinaud, C., Louarn, M., \& Conan, R. 2007, Applied optics,
  46, 6176

\bibitem[{Korkiakoski {et~al.}(2008)Korkiakoski, Vérinaud, \& Louarn}]{4k}
Korkiakoski, V., Vérinaud, C., \& Louarn, M.~L. 2008, in Adaptive Optics
  Systems, ed. N.~Hubin, C.~E. Max, \& P.~L. Wizinowich, Vol. 7015,
  International Society for Optics and Photonics (SPIE), 1422 -- 1431

\bibitem[{Meimon {et~al.}(2015)Meimon, Petit, \& Fusco}]{meimon}
Meimon, S., Petit, C., \& Fusco, T. 2015, {Optics Express}, 23, 27134

\bibitem[{Neichel {et~al.}(2016)Neichel, Fusco, Sauvage, Correia, Dohlen,
  El-Hadi, Blanco, Schwartz, Clarke, Thatte, Tecza, Paufique, Vernet,
  le~Louarn, Hammersley, Gach, Pascal, Vola, Petit, Conan, Carlotti,
  V{\'e}rinaud, Schnetler, Bryson, Morris, Myers, Hugot, Gallie, \&
  Henry}]{harmoni}
Neichel, B., Fusco, T., Sauvage, J.-F., {et~al.} 2016, in Astronomical
  Telescopes + Instrumentation

\bibitem[{Ragazzoni(1996)}]{raga}
Ragazzoni, R. 1996, Journal of Modern Optics, 43, 289

\bibitem[{Rigaut {et~al.}(1998)Rigaut, Veran, \& Lai}]{rigaut98}
Rigaut, F.~J., Veran, J.-P., \& Lai, O. 1998, in Adaptive Optical System
  Technologies, ed. D.~Bonaccini \& R.~K. Tyson, Vol. 3353, International
  Society for Optics and Photonics (SPIE), 1038 -- 1048

\bibitem[{Roddier(1981)}]{roddier81}
Roddier, F. 1981

\bibitem[{V\'{e}ran \& Herriot(2000)}]{Veran:00}
V\'{e}ran, J.-P. \& Herriot, G. 2000, J. Opt. Soc. Am. A, 17, 1430

\bibitem[{{V{\'e}rinaud}(2004)}]{verinaud}
{V{\'e}rinaud}, C. 2004, Optics Communications, 233, 27

\bibitem[{{Vigan} {et~al.}(2019){Vigan}, {N'Diaye}, {Dohlen}, {Sauvage},
  {Milli}, {Zins}, {Petit}, {Wahhaj}, {Cantalloube}, {Caillat}, {Costille}, {Le
  Merrer}, {Carlotti}, {Beuzit}, \& {Mouillet}}]{arthurZelda}
{Vigan}, A., {N'Diaye}, M., {Dohlen}, K., {et~al.} 2019, \aap, 629, A11

\end{thebibliography}

\end{document}